\documentstyle[aps,multicol,epsf,epsfig,axodraw]{revtex}

\def\ruleleft{\vspace{-2.5\baselineskip}\begin{multicols}{2}\ \linebreak\vspace{-\baselineskip}\hrulefill\raisebox
{0.84mm}{$\!\rfloor$}\[\]\end{multicols}\vspace{-1.5\baselineskip}}
\def\ruleright{\vspace{-1.5\baselineskip}\begin{multicols}{2}\ \linebreak\raisebox
{-2.45mm}{$\lceil\!$}\hrulefill\end{multicols}\vspace{-\baselineskip}}
\begin{document}

\title{Rod-like Polyelectrolytes in Presence of Monovalent Salt}
\author{Paulo S. Kuhn, Yan Levin\footnote{Corresponding author; e-mail: levin@if.ufrgs.br} \, , and Marcia C. Barbosa}
\address{
Instituto de F\'{\i}sica, Universidade Federal
do Rio Grande do Sul\\ Caixa Postal 15051, CEP 91501-970
, Porto Alegre, RS, Brazil}
\maketitle
\begin{abstract}
We investigate the properties of rigid polyelectrolyte solutions in presence of monovalent salt. The free energy within the Debye-H\"uckel-Bjerrum (DHBj) theory [M. E. Fisher and Y. Levin, {\it Phys. Rev. Lett.} {\bf 71}, 3826 (1993)] is constructed. It is found that at thermodynamic equilibrium the polyelectrolyte solution consists of clusters composed of one polyion and various counterions. The distribution of the cluster densities is determined by finding the minimum of the Helmholtz free energy. The osmotic pressure and the average charge of the cluster are found and their dependence on Manning parameter $\xi$ is elucidated. A good agreement with the experimental results is obtained.

\end{abstract}

\bigskip

\begin{multicols}{2}
\section*{\bf 1. Introduction}

Over the last three decades polyelectrolyte solutions have found a number of practical applications ranging from super-absorbents and hair conditioners to water treatment. Many biologically important molecules such as DNA, are also polyelectrolytes. Not withstanding their practical importance our theoretical understanding of the behavior of these complex molecules is still quite rudimentary \cite{Man1}. The fundamental problem which makes the study of polyelectrolytes so much more difficult than that of regular polymers is the long-ranged nature of the Coulomb force. The scaling theories, which have proven so useful for simple polymers have, so far, failed in the case of polyelectrolyte solutions. Even the mean-field theories are extremely hard to construct, as one tries to take a realistic account of the long-ranged electrostatic interactions.

In general a polyion is a polymer some of whose monomers are ionized (all of the ionized monomers have the same sign of charge). In most practical applications the polyions are dissolved in some solvent, usually water, causing a strong interaction between the charged monomers and the counterions. Besides this, already very complex interaction, in the case of a flexible chain one must also take into account the conformational degrees of freedom of the polymer, as well as the interactions between the macromolecules inside the solution. The full problem is extremely difficult to study, however, some simulations have been attempted \cite{Ste}. A somewhat simpler problem, an answer to which we shall attempt to elucidate in this paper, is the properties of rigid polyelectrolytes. An example of rigid polyelectrolyte is a solution of DNA segments. A major simplification resulting from restricting our attention to rigid polyions is that these can be modeled as cylinders, thus allowing us to bypass the complication of
 taking a full account of the conformational degrees of freedom. At high densities the rigid molecules tend to align forming a smectic phase. The periodic structure of the smetic phase allows us a major simplification of studying {\it one} polyion inside a Wigner-Seitz cell, since all of the cells are identical \cite{Fuoss}. At low volume fractions, when the solution is isotropic, the Wigner-Seitz picture is no longer valid, and a new approach must be found. On the other hand, it is well known that at infinite dilution the polyelectrolyte solutions obey the Manning limiting laws \cite{Man2}. Although derived in a somewhat {\it ad hoc} manner, the limiting laws have proven extremely successful and provide a ``boundary condition" that any theory of polyelectrolytes must satisfy. Recently we have extended the Debye-H\"uckel-Bjerrum (DHBj) theory \cite{FisLev} to rigid polyelectrolyte solutions and found that at infinite dilution it, indeed, reduces to the Manning limiting laws \cite{LevBarb}. However, consideri
ng the higher density corrections, we have found that the limiting laws should apply only at extremely low densities, since the corrections scale as $1/\ln \rho$, where $\rho$ is the density of polyions \cite{LevBarb}. What can then account for the success of limiting laws at densities which are not so small? In this paper we shall extend the DHBj theory to realistic concentrations encountered in most applications.

DHBj theory has proven to be successful in a variety of systems whose dominant interactions are due to the long-ranged Coulomb potential \cite{LLF}. The fundamental idea behind the theory is that the nonlinearities which are omitted in the process of linearization of the Poisson-Boltzmann equation can be re-introduced into the theory through a thermodynamic assumption that the oppositely charged particles can associate, forming clusters. The density of these clusters will, then, be determined by the condition that at equilibrium the total free energy of the system must be minimum. In this sense, the DHBj theory provides a kind of variational approximation to the complete, yet unknown, theory. By virtue of being linear, DHBj theory bypasses the internal inconsistencies which complicate many of the nonlinear theories of ionic solutions \cite{Ons1}.

In the case of symmetric electrolytes, the clusters correspond to the dipolar pairs formed when two oppositely charged ions come into a close contact \cite{Bj}. These are exactly the kind of configurations which become ``undervalued" when one linearizes the Boltzmann factor. For the polyelectrolyte, the polyions of which can carry a charge which can be many thousands units (unit=electron charge), there can be a large variety of clusters. In general we expect that there will be clusters consisting of one polyion and one counterion, one polyion and two counterions, one polyion and three counterions, etc. The successful theory should allow us to calculate the full distribution of cluster densities. In the previous work we have demonstrated that in the limit of infinite dilution only one characteristic cluster size is thermodynamically stable \cite{LevBarb}. This is the same as was previously postulated by Manning \cite{Man2}. However, as one moves away from infinite dilution, one must consider the full distribu
tion of cluster sizes \cite{LevBarb}, and this will be the objective of this paper.

\section*{\bf 2. The model}

We shall first define the Primitive Model of Polyelectrolyte (PMP) \cite{LevBarb}. Our system will consist of long cylindric polyions, the spherical counterions, and salt, inside a volume $V$. The polyions, of length $L$ and density $\rho_p$, have diameter $a_p$, and charge $-Zq$ uniformly distributed along the length of the cylinder. The distance between charged groups is $b \equiv L/Z$. The counterions will be modeled as rigid spheres of diameter $a_c$ and charge $+q$ located at the center. The overall charge neutrality requires that the density of counterions be $Z \rho_p$. In addition to the polyions and the counterions we shall allow for the presence of salt of density $\rho_{salt}$. In this work we shall consider a simple monovalent salt and, in particular, treat the positive ions of salt as identical to the counterions. We shall, therefore, consider only two species of small ions, the ones with charge $+q$, which we shall denote {\it counterions} indiscriminantly of whether they are derived from the p
olymer or the salt, and the {\it coions} with charge $-q$, derived from dissociation of salt molecules. The solvent will be modeled as a continuum medium of dielectric constant $D$. The distance of closest approach, $a$, between a polyion and a counterion is $a \equiv (a_p+a_c)/2$. This represents an ``exclusion cylinder" due to the hard-core repulsion (see Fig.$1$).

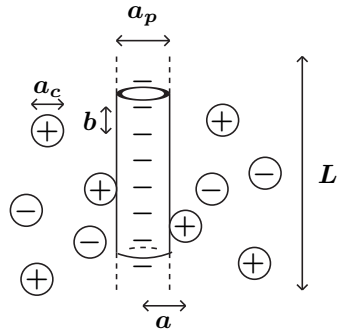
\begin{figure}[h]

\begin{center}
\begin{picture}(70,70)(0,0)
\Line(20,20)(20,50)
\Line(30,20)(30,50)
\DashLine(20,13)(20,20)1
\DashLine(20,50)(20,57)1
\DashLine(30,13)(30,20)1
\DashLine(30,50)(30,57)1
\Oval(25,50)(1.2,4.5)(0)
\DashCArc(25,6)(15,80,100)1
\CArc(25,33)(14,250,293)
\Text(25,17.5)[c]{\boldmath$-$}
\Text(25,22.5)[c]{\boldmath$-$}
\Text(25,27.5)[c]{\boldmath$-$}
\Text(25,32.5)[c]{\boldmath$-$}
\Text(25,37.5)[c]{\boldmath$-$}
\Text(25,42.5)[c]{\boldmath$-$}
\Text(25,47.5)[c]{\boldmath$-$}
\Text(25,52.5)[c]{\boldmath$-$}
\BCirc(7,43)3
\BCirc(5,15)3
\BCirc(15,22)3
\BCirc(3,28)3
\BCirc(17,32)3
\BCirc(33,25)3
\BCirc(48,35)3
\BCirc(38,32)3
\BCirc(46,18)3
\BCirc(40,45)3
\Text(5,15)[c]{\boldmath$+$}
\Text(15,22)[c]{\boldmath$-$}
\Text(3,28)[c]{\boldmath$-$}
\Text(17,32)[c]{\boldmath$+$}
\Text(7,43)[c]{\boldmath$+$}
\Text(33,25)[c]{\boldmath$+$}
\Text(48,35)[c]{\boldmath$-$}
\Text(38,32)[c]{\boldmath$-$}
\Text(46,18)[c]{\boldmath$+$}
\Text(40,45)[c]{\boldmath$+$}
\Line(4,48)(10,48)
\Line(4,48)(5,49)
\Line(4,48)(5,47)
\Line(10,48)(9,49)
\Line(10,48)(9,47)
\Text(7,51)[c]{\boldmath$a_c$}
\Line(55,13)(55,57)
\Line(55,13)(54,14)
\Line(55,13)(56,14)
\Line(55,57)(54,56)
\Line(55,57)(56,56)
\Text(60,35)[c]{\boldmath$L$}
\Line(18,42.5)(18,47.5)
\Line(18,42.5)(17,43.5)
\Line(18,42.5)(19,43.5)
\Line(18,47.5)(17,46.5)
\Line(18,47.5)(19,46.5)
\Text(15,45)[c]{\boldmath$b$}
\Line(20,60)(30,60)
\Line(20,60)(21,61)
\Line(20,60)(21,59)
\Line(30,60)(29,61)
\Line(30,60)(29,59)
\Text(25,65)[c]{\boldmath$a_p$}
\Line(25,10)(33,10)
\Line(25,10)(26,11)
\Line(25,10)(26,9)
\Line(33,10)(32,11)
\Line(33,10)(32,9)
\Text(29,7)[c]{\boldmath$a$}
\end{picture}
\end{center}
\vspace*{0.5cm}
\begin{minipage}{0.48\textwidth}
\caption{A polyion of (cylindric) diameter $a_p$ and length $L \gg a$ surrounded by spherical counterions and coions of diameter $a_c$. The charge spacing is $b \equiv L/Z$, and the radius of the exclusion cylinder is $a \equiv (a_p + a_c)/2$.}
\label{Fig.1}
\end{minipage}
\end{figure}

The strong electrostatic attraction between the polyions and the counterions will result in some counterions becoming associated with the polyion molecules, producing clusters consisting of {\it one} polyion molecule and $n$ counterions. Just as in the process of micellization, that is commonly observed in amphiphilic systems, we expect that there will be a distribution of cluster sizes. Thus, inside the polyelectrolyte solution, we shall encounter some free polyions with no attached counterions. We denote these zero-clusters of density $\rho_0$. We shall also encounter clusters consisting of {\it one} polyion and {\it one} counterion, {\it one} polyion and {\it two} counterions, etc. The density of $n$-clusters is $\rho_n$, with $n$ ranging from $1$ to $Z$. Since not all of the counterions condense onto the polyions, for any non-zero temperature some free, unassociated, counterions remain in the solution. We denote the density of unassociated counterions $\rho_+$. The conservation of the total number of par
ticles leads to two equations, namely

\begin{equation}
\label{m1}
\rho_p = \sum^Z_{n=0} \rho_n \, ,
\end{equation}

\noindent
and

\begin{equation}
\label{m2}
\rho_+ = \rho_{salt} + Z \rho_p - \sum^Z_{n=0} n \rho_n \, .
\end{equation}

\noindent
Since the coions do not participate in the association, their density remains unchanged, $\rho_-=\rho_{salt}$.

The goal of the theory is, therefore, to determine the distribution of cluster sizes. The complete thermodynamic information about the system is contained in the Helmholtz free energy. The condition that the free energy must be minimum will allow us to determine the distribution of cluster densities. Once the distribution is ascertained, all the thermodynamic functions of the system can be found through the appropriate operations on the free energy. For example, the pressure inside the polyelectrolyte solution is a Legendre transform of the Helmholtz free energy density $f=-F/V$,

\begin{equation}
\label{m2a}
p(T,\{\rho_t\}) = f(T,\{\rho_t\}) + \sum \mu_t \rho_t \, ,
\end{equation}

\noindent  
where the chemical potential of a specie of type $t$ (clusters, bare polyions, counterions, and coions) is $\mu_t=-\partial f/ \partial \rho_t$.

Unfortunately, there does not exist a way of calculating the free energy exactly. We shall, therefore, attempt to construct the approximate free energy as a sum of the most relevant contributions. These can be divided into an electrostatic and an entropic ones. The electrostatic contribution arises as the result of the polyion-counterion interaction, polyion-polyion interaction, and the counterion-coion interaction. The entropic contribution is the result of mixing of the various species \cite{LevBarb}.

\section*{\bf 3. The polyion-counterion interaction}

We calculate the polyion-counterion contribution to the free energy in the framework of the Debye-H\"uckel (DH) theory \cite{DH}. Consider a $n$-cluster fixed at the origin. We shall assume that as a counterion condenses onto a polyion it neutralizes one of its charged groups. Thus, the effective charge per unit length of a $n$-cluster is $\sigma_n=\sigma_0 (Z-n)/Z$, where $\sigma_0$ is the ``bare" charge per length, $\sigma_0 \equiv -Zq/L=-q/b$. The electrostatic potential around the cluster satisfies the Poisson equation,

\begin{equation}
\label{e9}
\nabla^2 \Phi^{(n)} = - \frac {4 \pi \rho_q} {D} \, .
\end{equation} 

\noindent
As a closure, we shall assume that the charge distribution is
\end{multicols}
\ruleleft
\medskip

\begin{eqnarray}
\label{e10}
\rho_q & = & \frac {\sigma_n} {2 \pi} \frac {\delta(r)} {r} \, , \,\,\,\, r<a \, , \\
& = & - \sum^Z_{n=0} (Z-n) q \rho_n + q \rho_+ e^{ - \beta q \Phi^{(n)}(r)} - q \rho_- e^{ +\beta q \Phi^{(n)}(r) } \, , \,\, r>a \, ,
\end{eqnarray}

\noindent
\vspace{-0.5\baselineskip}

\ruleright
\begin{multicols}{2}
\noindent
where $\beta=(k_B T)^{-1}$.
Note that only the counterions and the coions are assumed to get polarized, while the polyions and clusters are too massive to be affected by the electrostatic fluctuations and contribute only to the neutralizing background. 

The next step in the DH theory is to linearize the exponential. With the help of eqs. (\ref{m1}) and (\ref{m2}), the charge density eq. ($6$) becomes $\rho_q=- \rho_1 \beta q^2 \Phi^{(n)}(r)$, where $\rho_1=\rho_++\rho_-$. The nonlinear Poisson-Boltzmann equation reduces to a linear Helmholtz equation,

\begin{eqnarray}
\label{e11a}
\nabla^2 \Phi^{(n)} & = & - \frac {2 \sigma_n} {D} \frac {\delta(r)} {r} \, , \,\,\,\, r<a \, , \\
& = & \kappa^2 \Phi^{(n)} \, , \,\,\,\, r>a \, ,
\end{eqnarray}

\noindent
where $(\kappa a)^2 \equiv 4 \pi \rho^*_1/T^*$ and the reduced density and temperature are respectively $\rho_t^*=\rho_t a^3$ and $T^*=Dak_BT/q^2$.

Although the electrostatic energy close to a polyion is not small, the linearization is still valid if the bare charge is replaced by a renormalized charge \cite{AlCh}, which is {\it exactly} the idea behind the formation of clusters. 

\begin{figure}[h]

\begin{center}
\epsfxsize=8.cm
\leavevmode\epsfbox{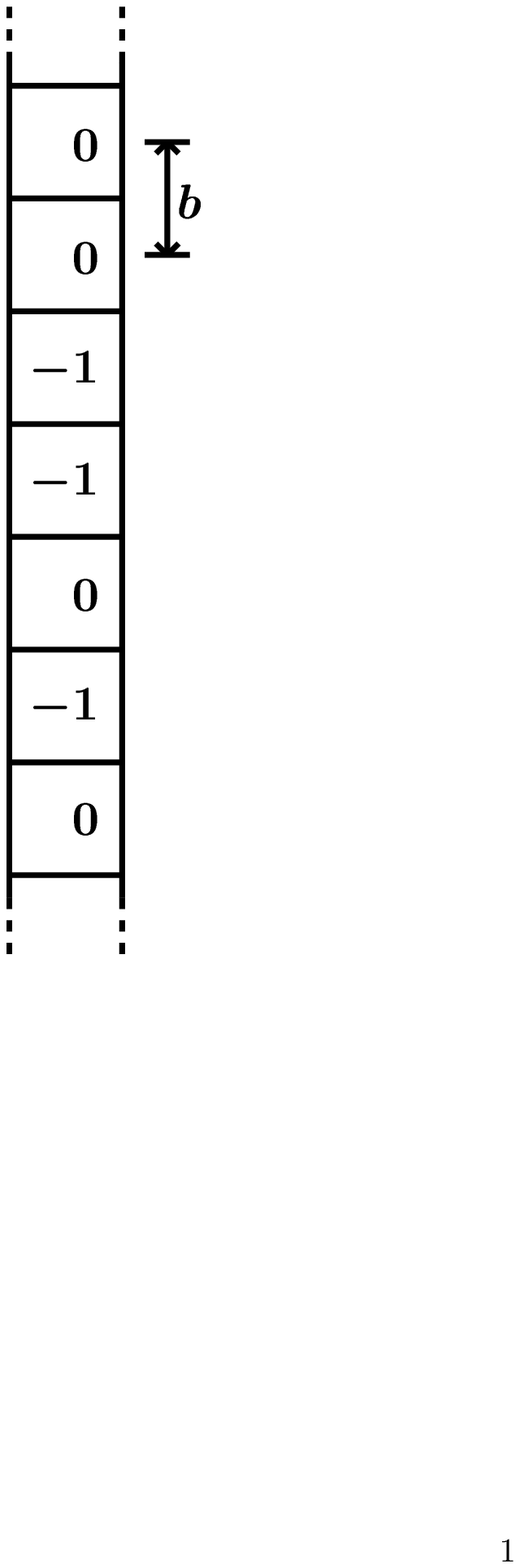}
\end{center}
\vspace*{0.5cm}
\begin{minipage}{0.48\textwidth}
\caption{Two infinite polyions at distance $p$ and with relative orientation $\theta$. The distance between the line elements $i$ and $j$ is $r^2=(i-j \cos \theta)^2+p^2+(j \sin \theta)^2$.}
\label{Fig.2}
\end{minipage}
\end{figure}
\noindent
The eqs. ($7$) and ($8$) reduce to a one-dimensional problem, since they are symmetric in the angle $\varphi$, and the $z$ dependence is suppressed due to the large extent of the polyion. The only cylindrical variable left is the radial distance $r$ from the polyion. The appropriate boundary conditions are: the vanishing of the potential for large values of $r$ and continuity of the potential and the electric field at $r=a$. With these conditions we find \cite{LevBarb}

\begin{eqnarray}
\label{e12}
& & \Phi_{in}^{(n)} = - \frac {2 \sigma_n} {D} \ln (r/a) + \frac {2 \sigma_n} {D} \Theta(\kappa a) K_0(\kappa a) \, , \,\,\,\, r<a \, , 
\\
& & \Phi_{out}^{(n)} = \frac {2 \sigma_n} {D} \Theta(\kappa a) K_0(\kappa r) \, , \,\,\,\, r>a \, ,
\end{eqnarray}

\noindent

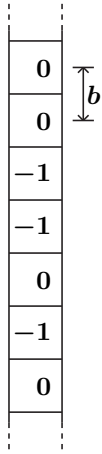
\begin{figure}[h]

\begin{center}
\epsfxsize=8.cm
\begin{picture}(10,90)(0,0)
\Line(0,8)(0,82)
\Line(10,8)(10,82)
\Line(0,10)(10,10)
\Line(0,20)(10,20)
\Line(0,30)(10,30)
\Line(0,40)(10,40)
\Line(0,50)(10,50)
\Line(0,60)(10,60)
\Line(0,70)(10,70)
\Line(0,80)(10,80)
\DashLine(0,3)(0,8)1
\DashLine(10,3)(10,8)1
\DashLine(0,82)(0,87)1
\DashLine(10,82)(10,87)1
\Text(8,15)[r]{\boldmath$0$}
\Text(8,25)[r]{\boldmath$-1$}
\Text(8,35)[r]{\boldmath$0$}
\Text(8,45)[r]{\boldmath$-1$}
\Text(8,55)[r]{\boldmath$-1$}
\Text(8,65)[r]{\boldmath$0$}
\Text(8,75)[r]{\boldmath$0$}
\Line(12,75)(16,75)
\Line(12,65)(16,65)
\Line(14,65)(14,75)
\Line(14,75)(13,74)
\Line(14,75)(15,74)
\Line(14,65)(13,66)
\Line(14,65)(15,66)
\Text(16,70)[c]{\boldmath$b$}
\end{picture}
\end{center}
\vspace*{0.5cm}
\begin{minipage}{0.48\textwidth}
\caption{A sample of charge distribution on a polyion. Each site contains $1$ or $0$ counterions associated to it, so the charge of the site is $0$ or $-q$, respectively.}
\vspace*{0.5cm}
\label{Fig.3}
\end{minipage}
\end{figure}

where

\begin{equation}
\label{e12a}
\Theta(x) \equiv \frac {1} {x K_1(x)} \, ,
\end{equation}

\noindent
and $K_n$ is the second-class modified Bessel function of order $n$. The geometric factor $\Theta(\kappa a)$ arises from the absence of screening inside $r<a$ \cite{LiLF}. If instead of a cylinder a polyion would be a line of charge, i.e., $a=0$, then the geometrical factor would become $\Theta(0)=1$.

The first term in eq. ($9$) is the potential of an infinite line of charge, while the second is the potential of the polyion due to the presence of other polyions and counterions. The electrostatic energy of a particular cluster of size $n$ may now be calculated through

\begin{equation}
\label{e13}
U^{(n)} = \frac {1} {2} \int \rho_q \, \Delta \Phi^{(n)} \, d^3r \, ,
\end{equation} 

\noindent
where $\Delta \Phi^{(n)} = \Phi^{(n)}_{in} + \frac {2 \sigma_n} {D} \ln (r/a)$, for $r<a$; $\Delta \Phi^{(n)} = \Phi^{(n)}_{out}$, for $r>a$. That is, we subtract the logarithmic potential produced by the line of charge, since it only contributes to the self energy of the cluster. The charge density is given by eqs. ($5$) and ($6$) in linearized form. Evaluating the integral, we find \cite{obs1}
\end{multicols}
\ruleleft
\medskip

\begin{eqnarray}
\label{e15}
\beta U^{(n)} = \frac {(Z-n)^2 (a/L)} {T^*} \frac {K_0(\kappa a)} {\kappa a K_1(\kappa a)} + \frac {(Z-n)^2 (a/L)} {2 T^*} \left( \frac {K_0^2(\kappa a)} {K_1^2(\kappa a)} - 1 \right) \, .
\end{eqnarray}

\noindent
\vspace{-0.5\baselineskip}

\ruleright
\begin{multicols}{2}

The electrostatic free energy is found through the Debye charging process, in which all the particles are charged simultaneously from $0$ to their final charge \cite{DH,Mar},

\begin{equation}
\label{e17}
f^{pc} = - \, \sum _{n=0}^Z \, \rho_n \, \int^{1}_{0} \, \frac {2 U^{(n)}(\lambda)} {\lambda} \, d \lambda \, .
\end{equation} 

\noindent

The free energy density due to the polyion-counterion interaction becomes
\end{multicols}
\ruleleft
\medskip

\begin{eqnarray}
\label{e19}
a^3 \beta f^{pc} & = & \sum^Z_{n=0} \rho^*_n (Z-n)^2 \frac {(a/L)} {T^*} \frac {1} {(\kappa a)^2} \left\{ 2 \ln \left[ \kappa a K_1(\kappa a) \right] - I(\kappa a) + \frac {(\kappa a)^2} {2} \right\} \, ,
\end{eqnarray}

\noindent
\vspace{-0.5\baselineskip}

\ruleright
\begin{multicols}{2}
with

\begin{equation}
\label{e20}
I(\kappa a) \equiv \int^{\kappa a}_{0} dx \frac {x K_0(x)^2} {K_1(x)^2} \, .
\end{equation}

\section*{\bf 4. The counterion-coion and the polyion-polyion interactions}

\subsection*{\bf 4.1 The counterion-coion interaction}

In the previous calculation of the polyion-counterion free energy we have neglected the correlational effects arising due to the counterion-counterion, counterion-coion, and the coion-coion interactions. To leading order the contribution to the excess free energy arising from the correlations in the charge density distribution can be calculated using the usual Debye-H\"uckel theory. To this end we consider the counterions and the coions moving in a neutralizing background produced by the free polyions and clusters. It is then a simple matter to demonstrate that the correlational free energy takes on the usual Debye-H\"uckel form \cite{DH,FisLev}

\begin{equation}
\label{e22}
a^3 \beta f^{cc} = \frac {1} {4 \pi} \left( \frac {a} {a_c} \right)^3 \left[ \ln \left[ 1+\kappa a_c \right] - \kappa a_c + \frac {(\kappa a_c)^2} {2} \right] \, .
\end{equation} 

\noindent

\subsection*{\bf 4.2 The polyion-polyion interaction}

The long-ranged Coulomb interaction between the macroions inside the polyelectrolyte solution will be screened by the mobile counterions and coions, producing a short-ranged effective potential. Following Onsager \cite{Ons2}, the energy of interaction between two {\it lines of charge} in an ionic sea is obtained by integrating the interaction of elements of charge, which is given by a Yukawa potential,

\begin{equation}
\label{e22a}
V_{mn} = \frac {\sigma_m \sigma_n} {D} \int^{+\infty}_{-\infty} di \int^{+\infty}_{-\infty} dj \frac {\exp \left( - \kappa r \right) } {r}.
\end{equation}

\noindent
Here $\sigma_m$ is the linear charge density of a polyion with $m$ counterions associated, $\kappa^2 = 4 \pi \rho_1 q^2 \beta/D$ is the inverse Debye length, eq. ($8$), $di$ is the line element of a polyion, and $r$ is the distance between the line elements,

\begin{equation}
\label{e22a1}
r^2 = (i-j \cos \theta)^2 + p^2 + (j \sin \theta)^2 \, .
\end{equation}

\noindent
Here $i$ and $j$ denote the distance of a line element to the center of each molecule, $0 \leq \theta \leq \pi$ is the relative angle of inclination between the lines, and $p$ is the perpendicular distance of separation between the molecules (see Fig.$2$).. Note that to simplify the calculations we have extended the limits of integration to infinity \cite{Ons2}. Changing the variables,

\begin{equation}
\label{e22a3}
s \equiv j \sin \theta \, , \,\,\,\,\, t \equiv i-j \cos \theta \, ,
\end{equation}

\noindent
we have

\begin{equation}
\label{e22a4}
r^2 = p^2 + s^2 + t^2 \, .
\end{equation}

\noindent

Hence, in terms of $s$ and $t$ the potential becomes 

\begin{equation}
\label{e22a5}
V_{mn} = \frac {\sigma_m \sigma_n} {D} \int \frac {ds \, dt} {\sin \theta} \frac {\exp \left[ - \kappa (p^2+s^2+t^2)^{1/2} \right] } {(p^2+s^2+t^2)^{1/2}} \, .
\end{equation}

\noindent

The integral can be performed easily by going to polar coordinates, and we find \cite{Ons2}

\begin{equation} 
\label{e22b}
V_{mn} = 2 \pi \sigma_m \sigma_n \frac {\exp \left( - \kappa p \right) } {\kappa D \sin \theta} \, .
\end{equation}

\noindent

So far in this section we have treated the molecules as lines of charge. A more realistic approximation is to model them as cylinders of radius $a$. As was mentioned following eq. ($10$), one can go from the line of charge to the cylinder by replacing $\sigma_n \rightarrow \sigma_n \Theta(\kappa a)$. In the case of spherical particles this transformation leads to the DLVO potential of interaction between two macroions \cite{LiLF}. We shall conjecture that the same procedure remains valid for cylinders at asymptotically large separations. With this geometrical factor, the pair potential of interaction between two clusters becomes 

\begin{equation}
\label{e22c}
V_{mn} = \frac {2 \pi \sigma_m \sigma_n}  {D \kappa \sin \theta} \frac {\exp \left( - \kappa p \right) } {(\kappa a)^2 K_1^2 (\kappa a)}.
\end{equation}

\noindent

It is important to note that this is an asymptotic potential, valid for $p \gg a$. When two polyions are in a close proximity of each other, the fluctuations in their condensed layers can become correlated, producing an effective attractive interaction \cite{Oos}. Though extremely interesting, this short-distance effect should not have any significant influence on the thermodynamic properties of the polyelectrolyte solution determined mostly by the counterions and their interactions with the polyions (see Fig.7).

Since the effective potential is short ranged, we can account for the polyion-polyion contribution to the free energy using a Van der Waals type of approach. To this end the polyion-polyion contribution is expressed as a second virial term averaged over the relative angle sustained by two molecules,
\end{multicols}
\ruleleft
\medskip

\begin{eqnarray}
\label{e22d}
f^{pp} & = & - \frac {1} {2} \sum_{m,n} \rho_m \rho_n \langle \int d^3r V_{m n} (r) \rangle_{\theta} \, , \nonumber \\
& = & - \frac {1} {2} \sum_{m,n} \rho_m \rho_n \langle 2 \, \int_{2a}^{+\infty} \, dp \, \int_{-\frac {L} {2} \sin \theta}^{+\frac {L} {2} \sin \theta} \, dy \, \int_{-\frac {L} {2}}^{+\frac {L} {2}} \, dz V_{mn}\, \rangle_{\theta} \, .
\end{eqnarray}

\noindent

The result is

\begin{equation}
\label{e22e}
\beta f^{pp} = \frac {-2 \pi a^3} {T^*} \frac {\exp \left( -2 \kappa a \right) } {(\kappa a)^4 K_1(\kappa a)^2} \sum_{m,n} (Z-m) (Z-n) \rho_m \rho_n \, .
\end{equation}

\noindent
Substituting (\ref{m2}) we obtain 

\begin{equation}
\label{e23}
a^3 \beta f^{pp} = \frac {-2 \pi (\rho^*_1-2 \rho_{salt}^*)^2} {T^*} \frac {\exp (-2 \kappa a) } {(\kappa a)^4 K_1(\kappa a)^2} \, .
\end{equation} 

\noindent


\begin{figure}[h]

\begin{center}
\epsfxsize=8.cm
\leavevmode\epsfbox[40 120 550 520]{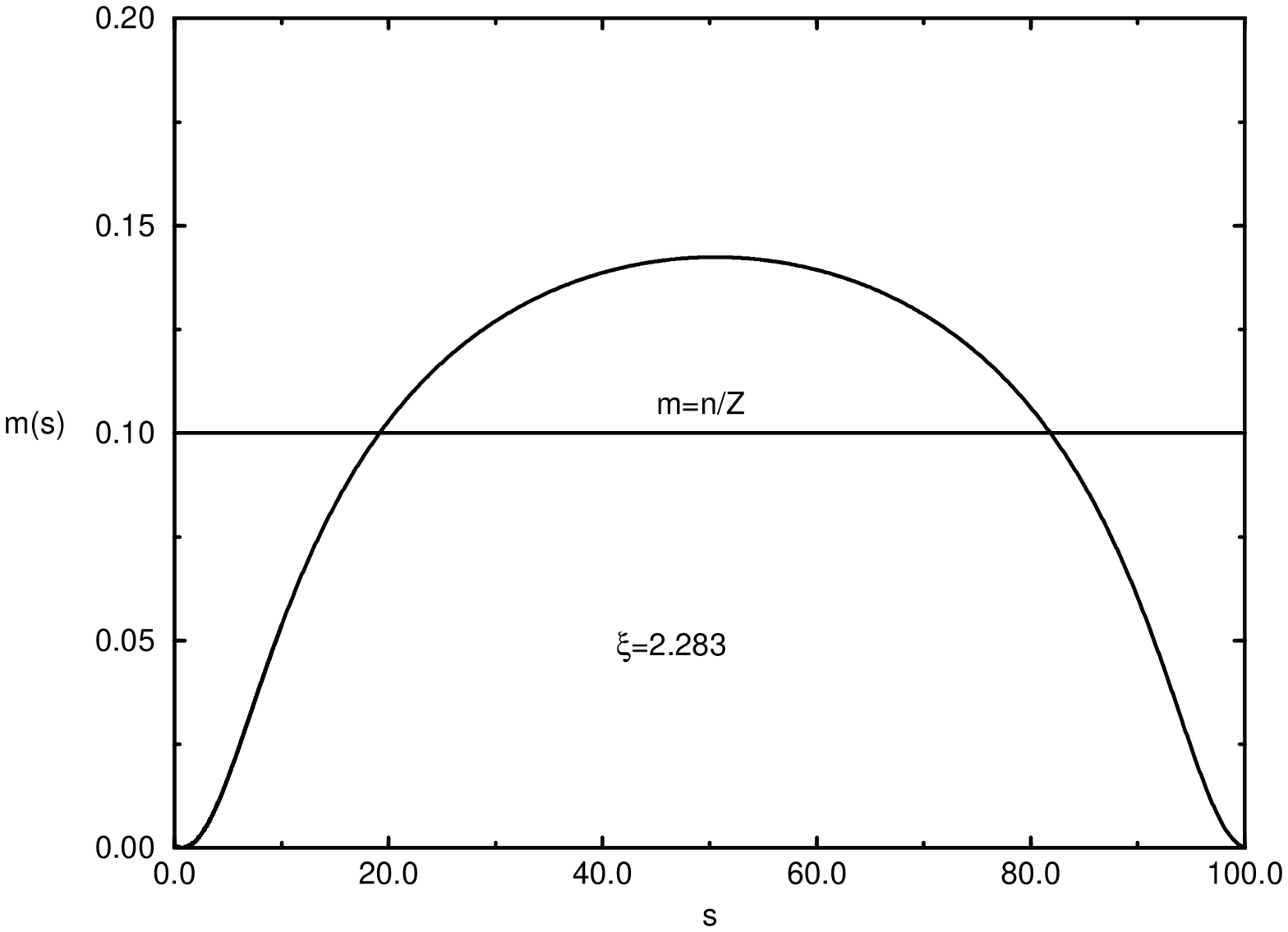}
\epsfxsize=8.cm
\leavevmode\epsfbox[40 120 550 520]{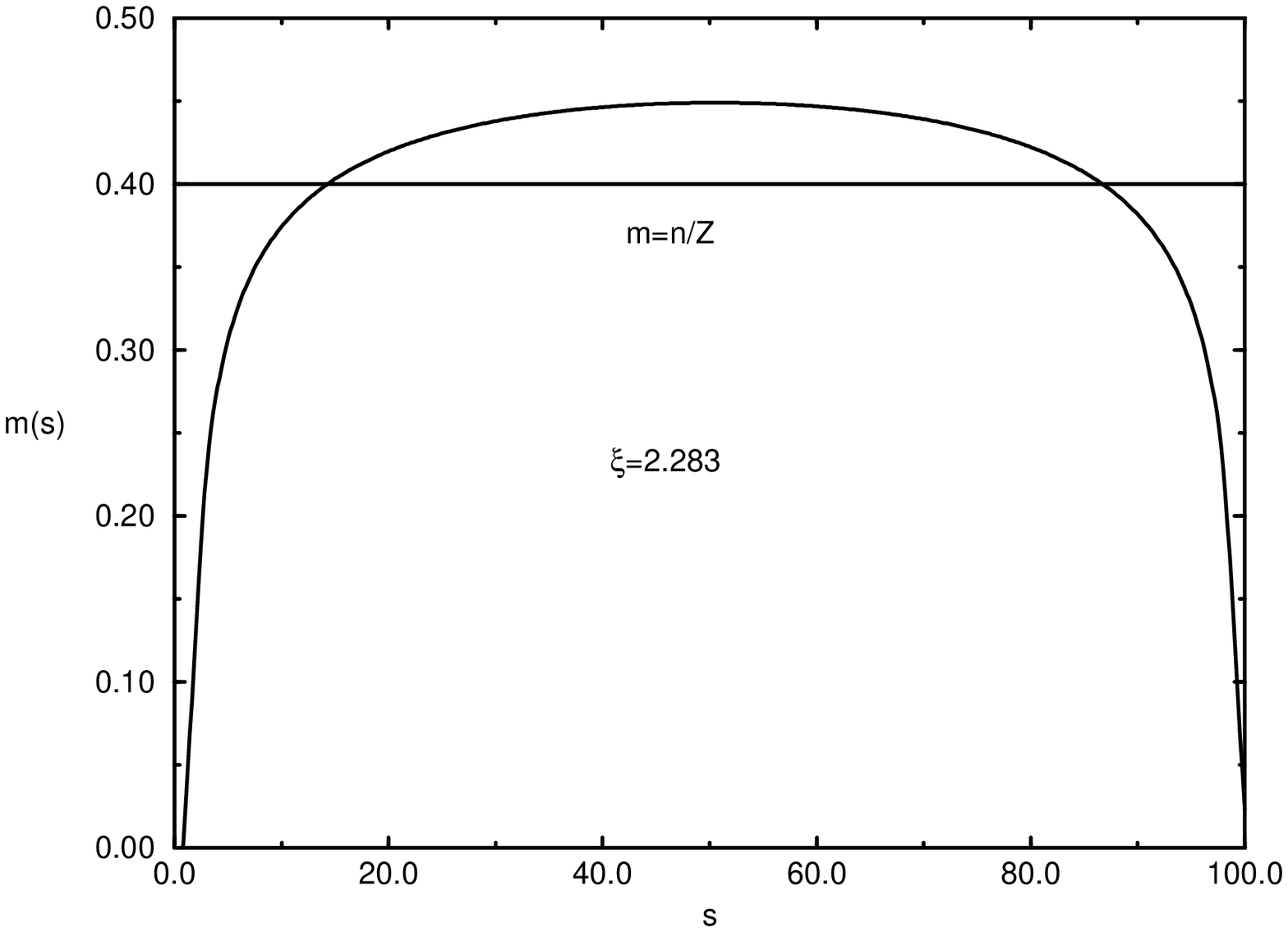}
\epsfxsize=8.cm
\leavevmode\epsfbox[40 120 550 520]{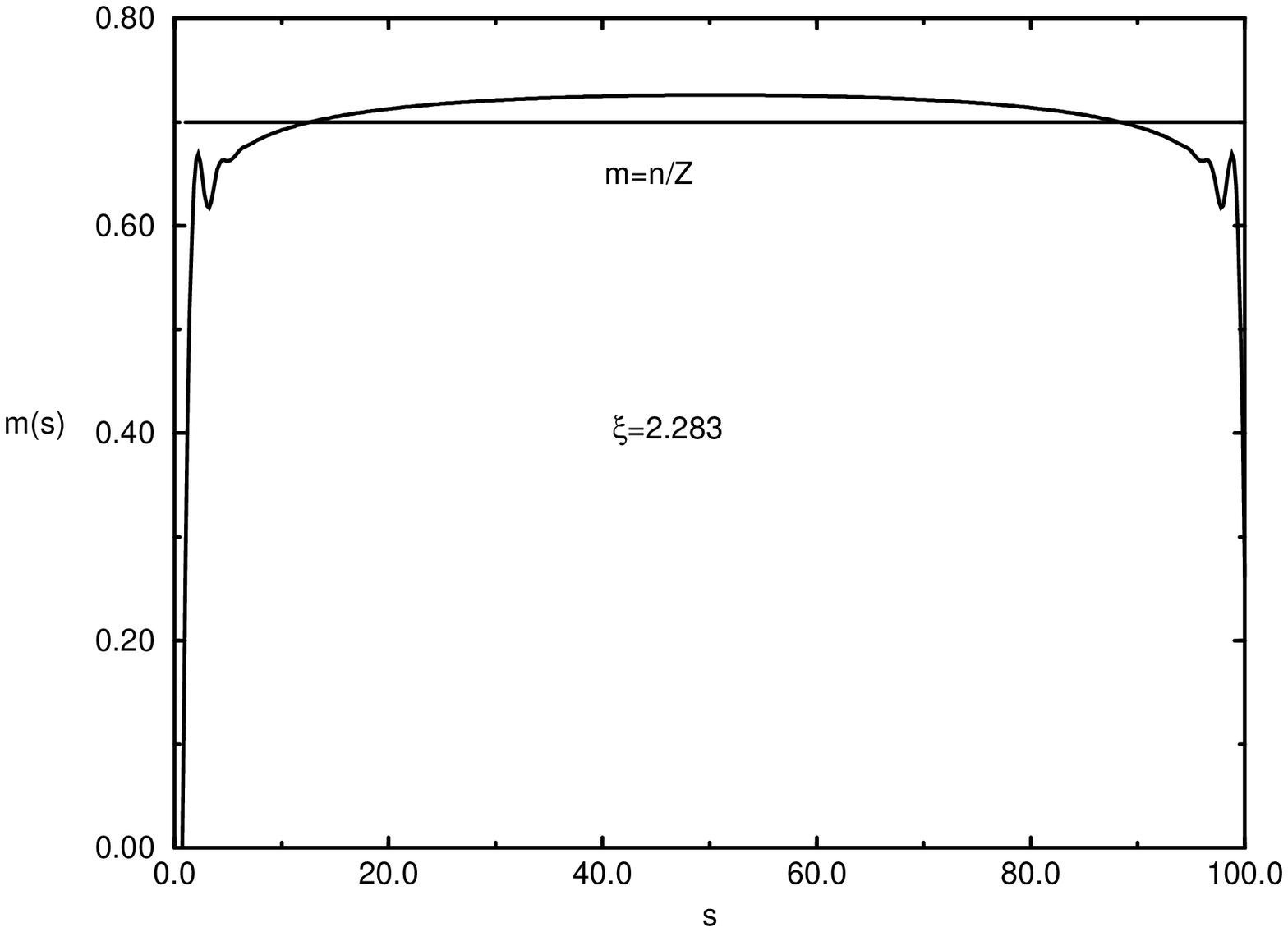}
\epsfxsize=8.cm
\leavevmode\epsfbox[40 120 550 520]{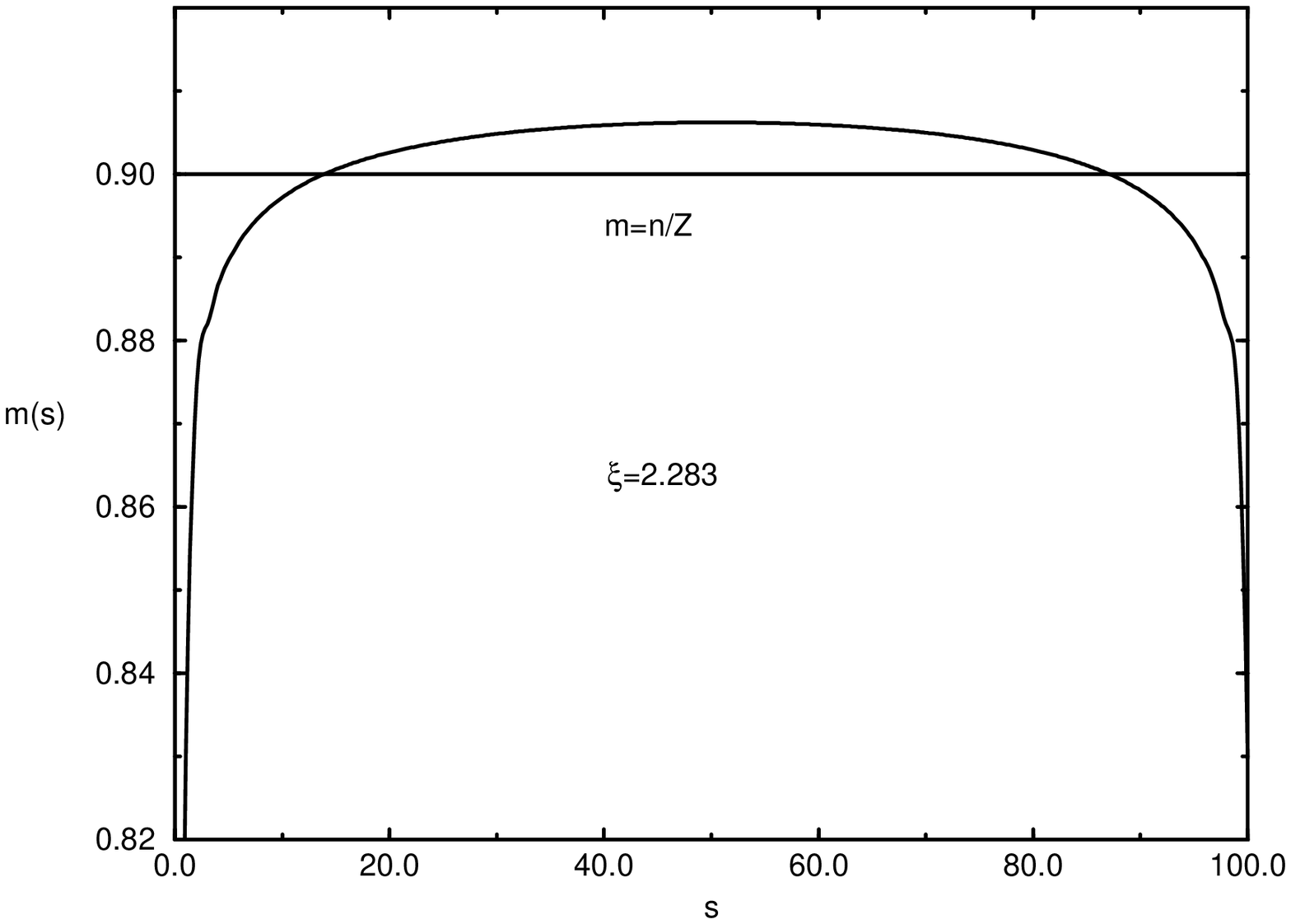}
\end{center}
\vspace*{0.5cm}
\caption{Distribution $\bar{m}(s)$ of associated counterions on a $n$-cluster for $Z=100$, and several values of $n$. (a) $n=10$; (b) $n=40$; (c) $n=70$; (d) $n=90$.}
\label{Fig.4}
\end{figure}
\vspace{0.5cm}

\begin{multicols}{2}

\section*{\bf 5. The mixing free energy}

The entropic contribution to the free energy can be calculated using the Flory theory \cite{Flo}. Thus the increase in free energy due to mixing of various species is the sum of their ideal free energies,

\begin{equation}
\label{e1}
\beta f^{ent} = \sum_t \left[ \rho_t - \rho_t \ln \left( \phi_t/\zeta_t \right) \right] \, .
\end{equation}

\noindent
The volume fractions are

\begin{eqnarray}
\label{e2}
& & \phi_n = \frac {\pi \rho^*_n} {4 (a/L)} \left(\frac {a_p} {a}\right)^2 + \frac {n \pi \rho^*_n} {6} \left(\frac {a_c} {a}\right)^3 \, , \\
& & \phi_+ = \frac {\pi \rho^*_+} {6} \left( \frac {a_c} {a} \right)^3 \, , \\
& & \phi_- = \frac {\pi \rho^*_-} {6} \left( \frac {a_c} {a} \right)^3 \, ,
\end{eqnarray}

\noindent
for the $n$-clusters, the counterions, and the coions respectively; $\zeta_t$ is the internal partition function of an isolated specie $t$. For structureless particles, $\zeta_+=\zeta_-=\zeta_0 \equiv 1$, while for clusters 

\begin{equation}
\label{e2a}
\zeta_n = \frac {\sum_{q(s)} \exp (- \beta H[q(s)])} {\exp (- \beta H[-q])} \, .
\end{equation}

\noindent

The Hamiltonian $H$ represents the electrostatic energy of interaction between the $n$ condensed counterions located at sites ${s_i}$, $i=1,...,n$, and the negatively charged monomers of the polyion (see Fig.$3$),

\begin{equation}
\label{e7a}
H[q(s)] = \frac {1} {2} \sum_{s_1 \not= s_2} \frac {q(s_1) q(s_2)} {D |r(s_1)-r(s_2)|} \, .
\end{equation}

\noindent


\vspace{0.5cm}
\begin{center}
\begin{table}[t]
\begin{minipage}{0.48\textwidth}
\caption{Free energy of association for $Z=100$ and $\xi=2.283$. $\bar{F}^R[\bar{m}(s)]$ is the free energy obtained using the variational counterion profile. $\bar{F}^R[m]$ is the free energy with a uniform profile $m=n/Z$.}
\begin{tabular}{c c c} \hline \hline \\
$n$ & $\bar{F}^R[\bar{m}(s)]$ & $\bar{F}^R[m]$ \\ \hline \hline
\medskip
$10$ & $-218.443899$ & $-214.144172$ \\
\hline
$20$ & $-399.827094$ & $-394.192426$ \\
\hline
$30$ & $-554.221516$ & $-548.635357$ \\
\hline
$40$ & $-683.959854$ & $-679.127270$ \\
\hline
$50$ & $-790.048546$ & $-786.298434$ \\
\hline
$60$ & $-872.900058$ & $-870.322928$ \\
\hline
$70$ & $-932.493432$ & $-931.026672$ \\
\hline
$80$ & $-968.353696$ & $-967.779398$ \\
\hline
$90$ & $-979.023957$ & $-978.926802$ \\
\hline \hline
\end{tabular}
\end{minipage}
\end{table}
\end{center}



We are labeling the $Z$ sites on the polyion with the index $s$, so $1 \leq s \leq Z$; $q(s)$ is the net charge on site $s$, which is $-q$ (no counterion associated) or $0$ (one counterion associated); $r(s)$ is the distance (in units of $b$) of site $s$ from one of the ends. We assume that the only effect of a condensed counterion is to neutralize the monomer charge. Furthermore, the electrostatic repulsion between the counterions will prevent that a site is occupied by more than one counterion. $H[-q]$ is the energy of the reference state in which no counterions are condensed, and each site has the charge $-q$.

At this point we need to evaluate the internal partition function. Unfortunately even this comparatively simple problem is very difficult to solve due to the long-ranged nature of the Coulomb force. In the absence of the exact solution we shall resort to a mean-field type of approach. We first define the free energy of association as $-\beta F^R = \ln \zeta_n$. It is easy to show that this energy is bounded from above,

\begin{equation}
\label{e7}
F^R < \bar{F}^R \equiv {F}^{R}_0 + \langle H-H_0 \rangle_0 - H[-q] \, ,
\end{equation} 

\noindent
where $H_0$ is an arbitrary Hamiltonian and $F^R_0$ is the free energy corresponding to $H_0$. This is a well known Gibbs-Bogoliubov bound on free energy. We shall choose $H_0$ to be of one-body form so as to facilitate the calculations,

\begin{equation}
\label{e7f}
H_0 [q(s)]= \sum_s \varphi(s) q(s) \, , \,\, q(s)=-q,0 \, ,
\end{equation}

\noindent
where $\varphi(s)$ is the effective position dependent mean field to be determined. The partition function corresponding to $H_0$ is
\end{multicols}
\ruleleft
\medskip

\begin{equation}
\label{e7b}
Z_0 = \prod_s \sum_{q(s)=-q,0} \exp \left[ -\beta \sum_s \varphi(s) q(s) \right] = \prod_s \left\{ \exp [+ \beta q \varphi(s)]+1 \right\} \, ,
\end{equation}

\noindent
\vspace{-0.5\baselineskip}

\ruleright
\begin{multicols}{2}
and the free energy,

\begin{equation}
\label{e7c}
F^R_0=- \frac {1} {\beta} \sum_s \ln \left\{ \exp [\beta q \varphi(s)] +1 \right\} \, .
\end{equation}

\noindent
The average charge on site $s$ is

\begin{equation}
\label{e7d}
p(s) \equiv \langle q(s) \rangle_0 = - \frac {q \exp \left[ \beta q \varphi(s)\right] } {\exp \left[ \beta \varphi(s) q \right] + 1}.
\end{equation}

\noindent
>From this we obtain

\begin{equation}
\label{e7e}
\exp \left[ \beta q \varphi(s) \right] = -\frac {p(s)} {p(s)+q} \, .
\end{equation}

\noindent


Therefore,
\end{multicols}
\ruleleft
\medskip

\begin{eqnarray}
\label{e8}
\bar{F}^R[m(s)] & = & \frac {1} {2} \frac {a} {T^*} \sum_{s_1,s_2} \frac {m(s_1) m(s_2) - m(s_1) - m(s_2)} {|r(s_1) - r(s_2)|} + \, \sum_s [1-m(s)] \ln \left[ 1-m(s) \right] \, + \, \sum_s m(s) \ln \left[ m(s) \right] \, ,
\end{eqnarray} 

\noindent

\begin{multicols}{2}
where $m(s)=\frac {p(s)} {q} +1$ is the average number of associated counterions on site $s$.

To find the optimal bound on free energy we shall minimize it with respect to $m(s)$, i.e., $\delta \bar{F}^R / \delta m(s) = 0$, under the constraint $\sum_s m(s) = n$. The resulting distribution $\bar{m}(s)$, which is obtained by numerically solving the minimization problem (see Fig. $4$), corresponds (within mean field theory) to the equilibrium density profile, i.e., $\langle q(s) \rangle /q +1 \approx \bar{m}(s)$. Without the knowledge of the exact solution we shall approximate the free energy of association by its optimal bound, $F^R \approx \bar{F}^R[\bar{m}(s)]$. In practice, however, even this simplified procedure will make finding the distribution of cluster sizes very slow, since for each cluster size $n$ we will need to find $\bar{m}(s)$ by numerically solving the minimization problem. Instead, we noted that outside the endcap regions, $\bar{m}(s)$ is more or less uniform and can be approximated by $\bar{m}(s) \approx m \equiv n/Z$ (see Fig.$4$). With this approximation the sums in eq. ($40$) can
 be carried out explicitly and we find ($\xi = \beta q^2/Db$ is the Manning parameter)

\begin{eqnarray}
\label{e8a}
\beta \bar{F}^R[m] & = & \xi \, (m^2 - 2 m) \, \left\{Z \, [ \psi(Z) - \psi(1)] - Z + 1 \right\} \nonumber \\
& + & Z \left[ (1-m) \ln \left( 1-m \right) \, + \, m \ln m \right] \, ,
\end{eqnarray}

\noindent
where $\psi(n)$ is the digamma function \cite{GR}. We have compared the free energy obtained using this uniform counterion profile with that obtained using the exact solution of the variational problem and found that they agree to within a few percent (see Table $1$). We shall, therefore, approximate the internal partition function of a $n$-cluster as $\zeta_n = \exp (- \beta \bar{F}^R[m])$ with $m=n/Z$. Inserting this into eq. ($28$) completes the calculation of the entropic contribution to the free energy inside the polyelectrolyte solution.
\end{multicols}


\begin{figure}[h]

\begin{center}
\epsfxsize=8.cm
\leavevmode\epsfbox[40 120 550 520]{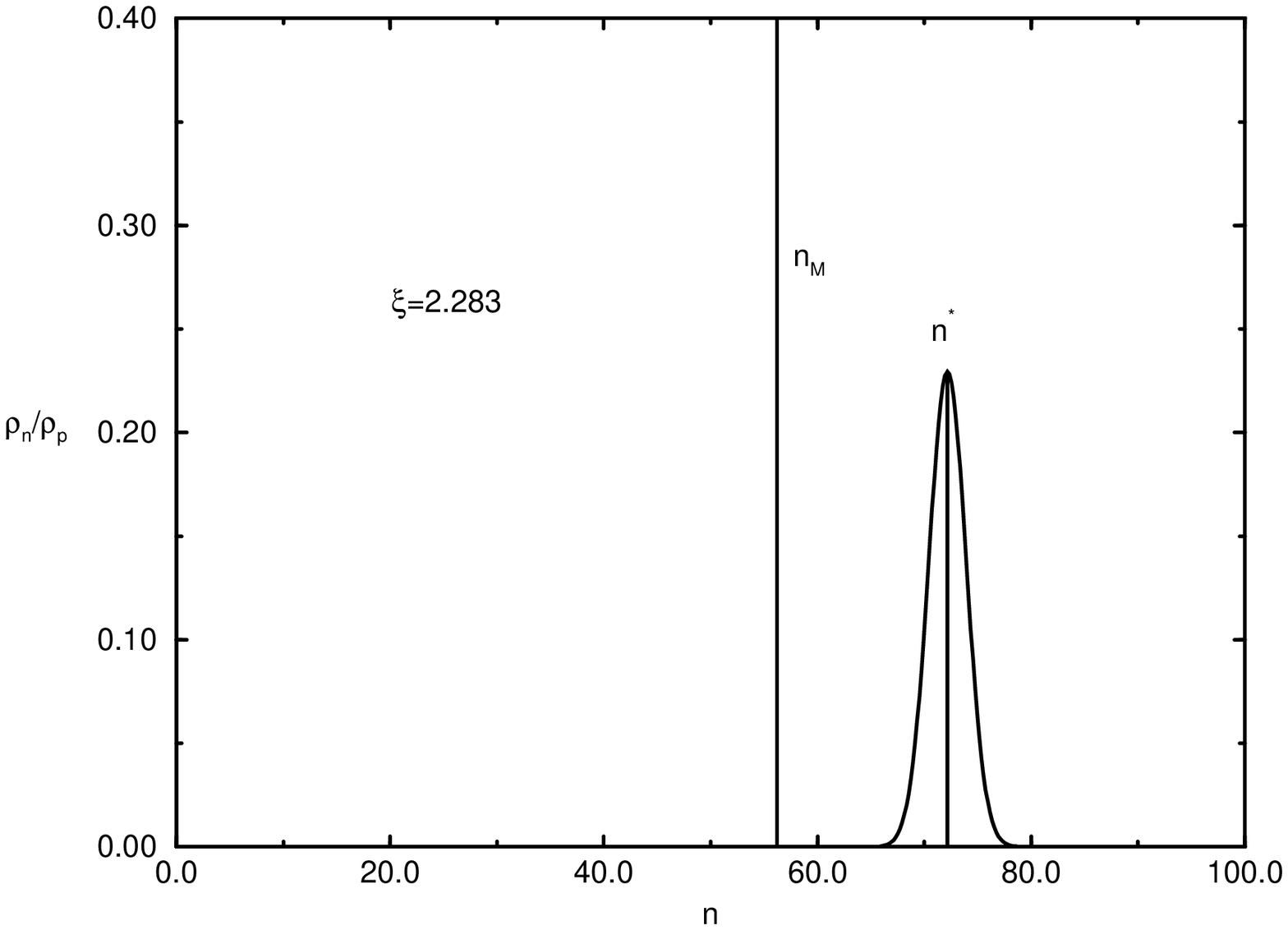}
\epsfxsize=8.cm
\leavevmode\epsfbox[40 120 550 520]{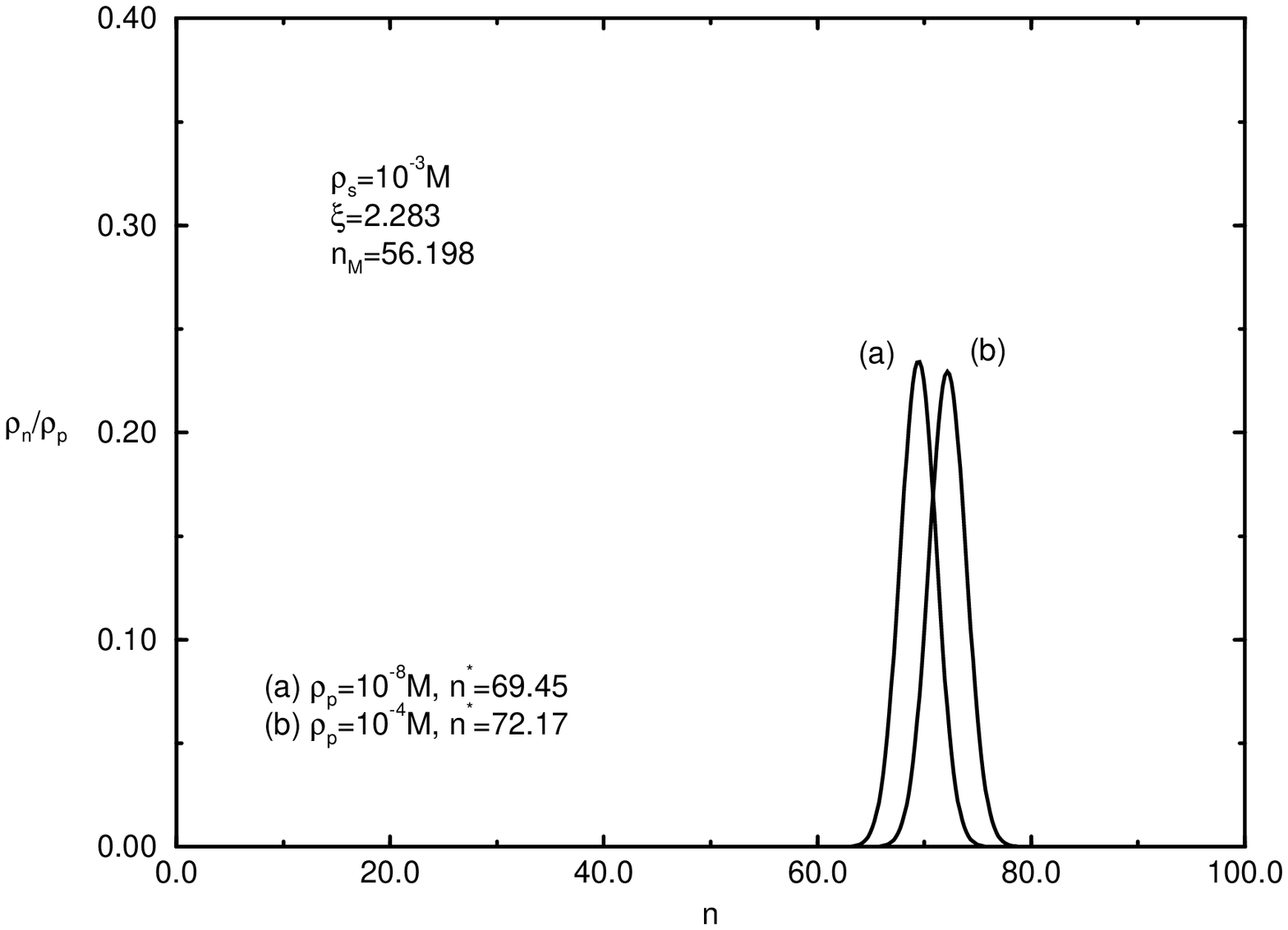}
\end{center}
\vspace*{0.5cm}
\caption{(a) Cluster size distribution $\rho_n$ for polyion density $\rho_p=10^{-4}$M, salt $\rho_{salt}=10^{-3}$M, and $Z=100$; the polyion and counterion diameters are $12$\AA \, and $9.3$\AA, respectively. The average number of associated counterions is $n^*=72.173$, while the Manning infinite-dilution result is $n_M=56.198$. (b) Cluster size distributions $\rho_n$ for two different densities of polyions. One should note the strong insensitivity of the distribution to the variation in density of polyions. Note changing scale.}
\end{figure}

\begin{multicols}{2}

\section*{\bf 6. The Helmholtz free energy}

The full Helmholtz free energy for the polyelectrolyte plus salt system is

\begin{equation}
\label{e24}
f = f^{ent} + f^{pc} + f^{cc} + f^{pp} \, .
\end{equation} 

\noindent

>From the free energy all the thermodynamic properties of the solution can, in principle, be found. Minimization of the free energy under the constraint of fixed number of particles, eqs. (\ref{m1}) and (\ref{m2}), reduces to the law of mass action, $\mu_0 + n \mu_+ = \mu_n$. The chemical potentials of all species are obtained from the free energy, $\mu_t = - \partial f/\partial \rho_t$, and in thermodynamic equilibrium we find that

\begin{equation}
\label{e25}
\phi_n = \phi_0 \, \phi_+^n \, \zeta_n \, \exp [\beta (\mu^{ex}_0 + n \mu^{ex}_+ - \mu^{ex}_n)] \, ,
\end{equation} 

\noindent
where the excess chemical potential is $\mu_t^{ex} = - \partial f^{ex}/\partial \rho_t$; $f^{ex}=f^{pc}+f^{cc}+f^{pp}$, and the $\phi$'s are the volume fractions defined earlier. The above expression ($43$) consists of $Z+1$ equations for the $Z+1$ densities involved, and its solution represents a formidable task. However, the numerical treatment of this problem is feasible, and employing an iterative algorithm we were able to find the distribution of cluster sizes (see Fig.$5$).
\end{multicols}

\begin{multicols}{2}
With the cluster densities determined the osmotic pressure may be calculated, and the results for several values of densities of salt are plotted in Fig.$6$. A good agreement with Manning's low-density limit and the experimental results of Alexandrowicz \cite{Alex} is obtained. The relative contribution of each term of the free energy can be seen in Fig.$7$. Note that the polyion-polyion contribution to the osmotic pressure is negative, what can be interpreted as an effective induced attraction between the polyions. The characteristic size of a cluster is $n^*=\sum_n n \rho_n /\rho_p$. In Fig.$8$ we demonstrate the dependence of the average charge of a cluster,

\begin{equation}
\label{e26}
\sigma_{eff} \equiv \sum_{n=0}^Z \frac {\sigma_n \rho_n} {\rho_p} \, = \sigma_0 \, \left( 1 - \frac {n^*} {Z} \right) \, ,
\end{equation} 

\noindent
on $1 / \xi$, and compare it to the infinite-dilution counterion condensation theory \cite{Man2}. We note that the sharp counterion condensation transition disappears at finite densities, replaced by a smooth crossover. Finally, in Fig.$9$ we demonstrate the relative insensitivity of the average cluster charge on the density of added salt.
\end{multicols}

\vspace{0.5cm}
\begin{figure}[h]

\begin{center}
\epsfxsize=8.cm
\leavevmode\epsfbox[40 120 550 520]{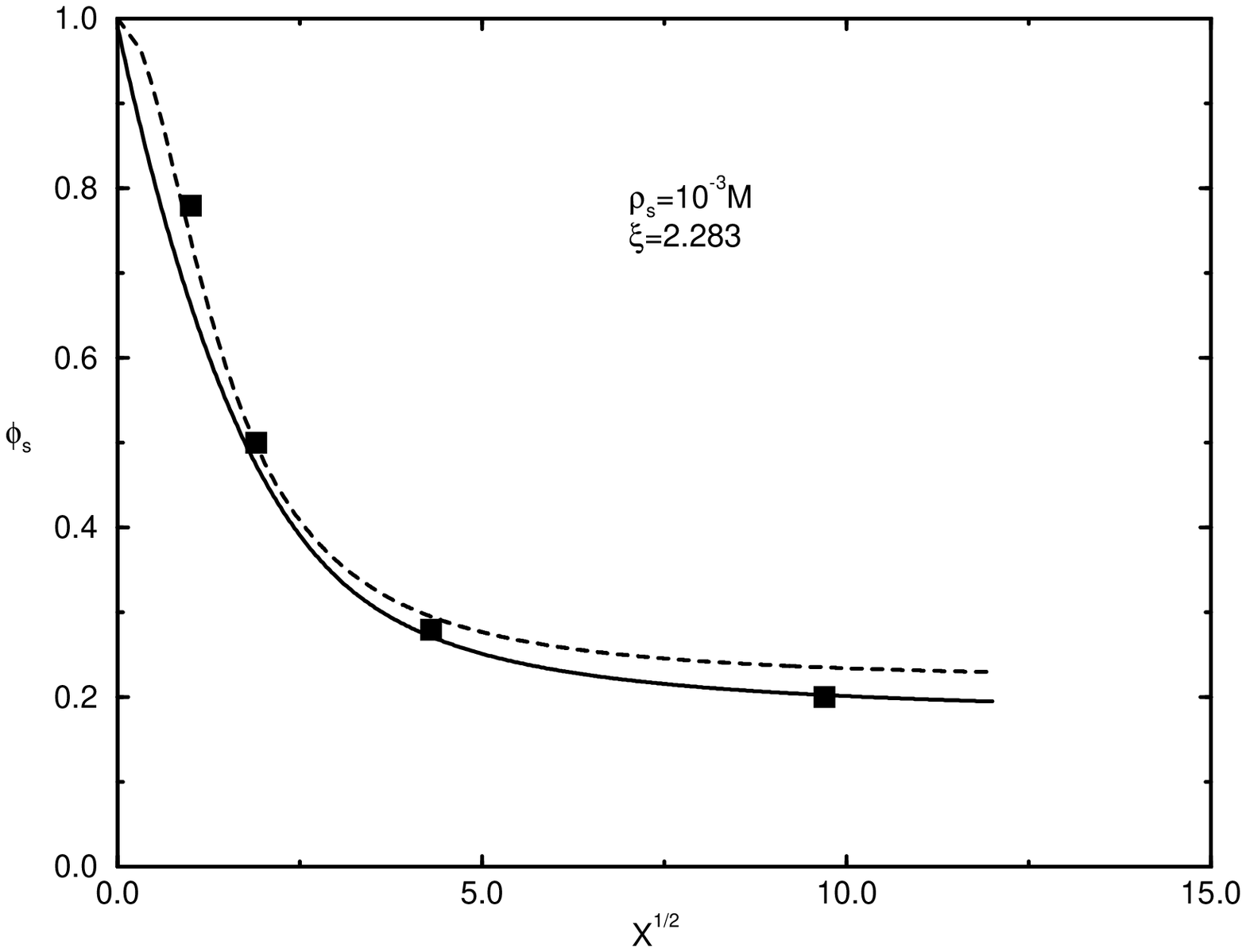}
\epsfxsize=8.cm
\leavevmode\epsfbox[40 120 550 520]{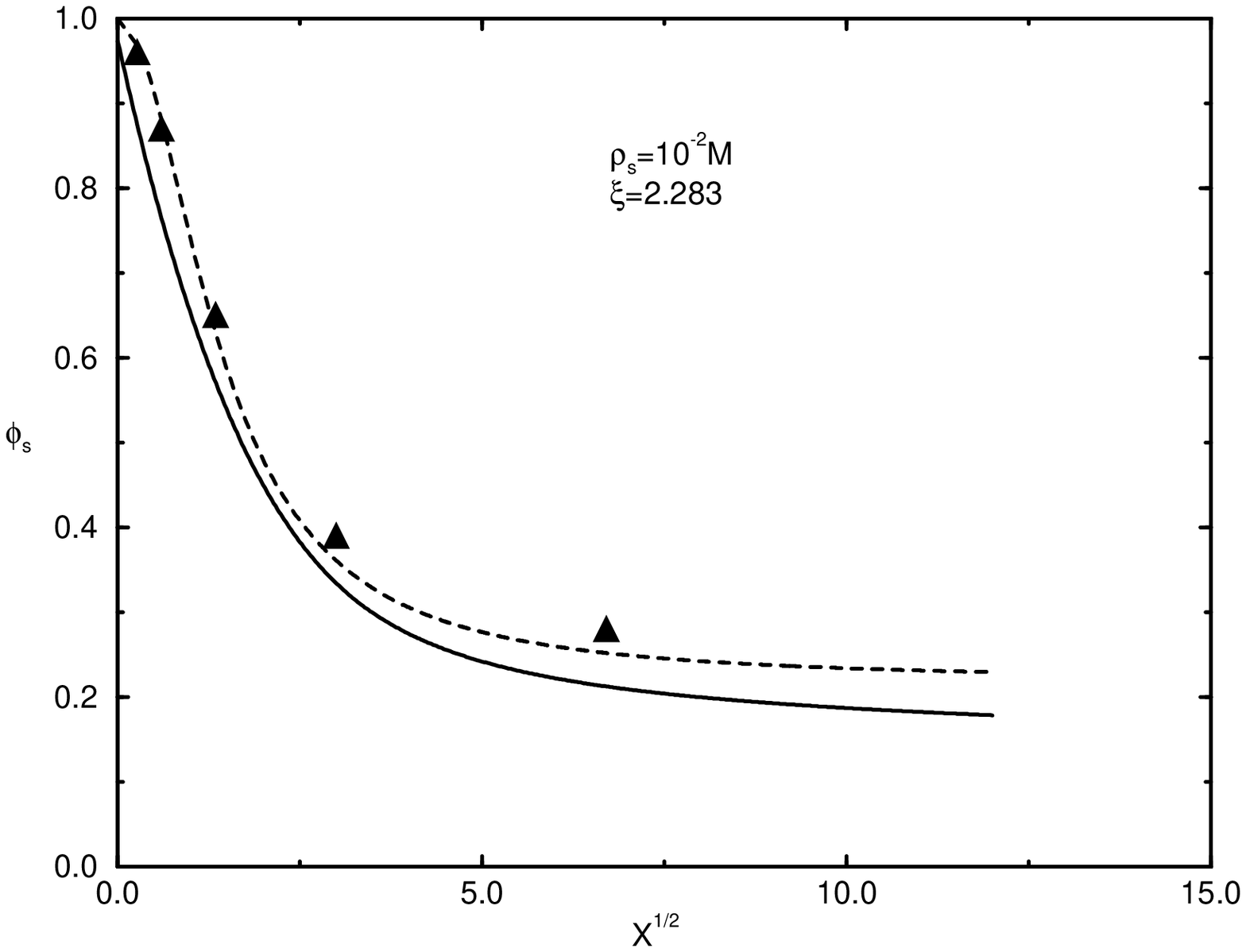}
\epsfxsize=8.cm
\leavevmode\epsfbox[40 120 550 520]{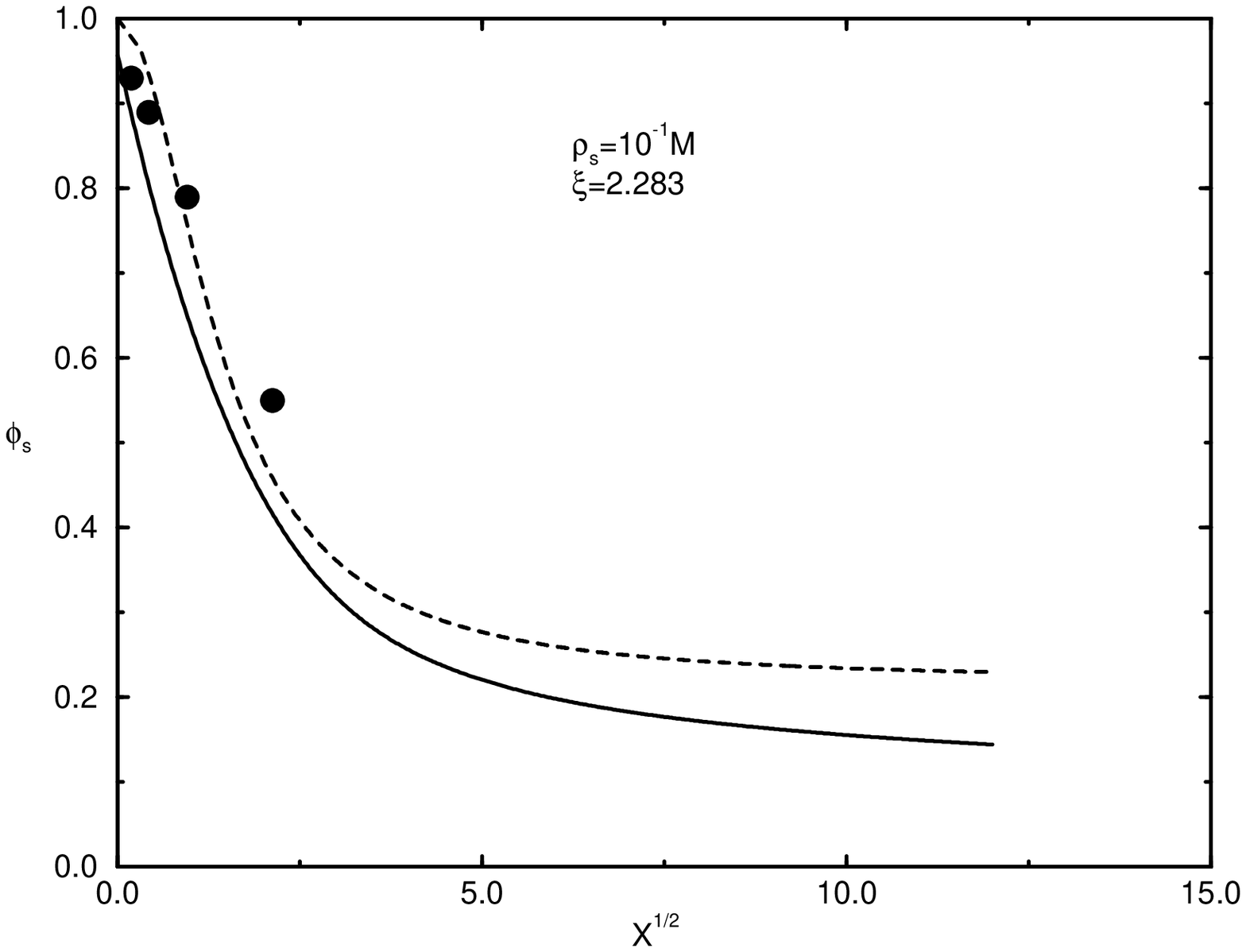}
\end{center}
\vspace*{0.5cm}
\caption{Osmotic pressure for a polyelectrolyte solution (solid line) for three different concentrations of salt: (a)$\rho_{salt}=10^{-3}$M; (b)$\rho_{salt}=10^{-2}$M; (c)$\rho_{salt}=10^{-1}$M. We have defined $X \equiv Z \rho_p / \rho_{salt}$. The characteristics of the polyions are: $Z=100$, $\xi=2.283$, $a_p=12$\AA, $a_c=9.3$\AA, $a=10.65$\AA. The dashed line is the infinite-dilution limit of Manning, while the squares, triangles and circles are the experimental results of Alexandrowicz (see text for the references).}
\label{Fig.6}
\end{figure}

\newpage

\begin{multicols}{2}

\section*{\bf 7. Conclusions}

We have constructed a DHBj theory for an isotropic polyelectrolyte solution. We find that, at equilibrium, the solution consists of free polyions, free counterions, and clusters composed of one polyion and $1 \leq n \leq Z$ counterions. The distribution of cluster sizes was calculated explicitly as was the osmotic pressure. A good agreement with the experimental data of Alexandrowicz was obtained. Finally, it is important to note that the sharp counterion condensation observed at $\xi = 1$ for infinite dilution, gets ``smoothed out" for finite densities. This is indeed what one might have expected based on the modern theory of phase transitions. It is well known that the higher-order phase transitions (second and above) are associated with a diverging length scale. In the ionic systems the length scale relevant to the counterion condensation phenomena is the Debye screening length. In the case of polyelectrolyte solutions, this length will stay finite for any non-zero temperature. In this respect the counter
ion condensation is very different from the Kosterlitz-Thouless (KT) metal-insulator transition observed in two-dimensional plasma \cite{KT}. The KT transition is a real thermodynamic phase transition at which the Debye length becomes infinite. It is important to stress this difference in view of the recent speculations that the counterion condensation is a KT-like phase transition \cite{KB}. Instead we should compare the counterion association with a micellar formation in amphiphilic systems.


\begin{figure}[h]

\begin{center}
\epsfxsize=8.cm
\leavevmode\epsfbox[40 120 550 520]{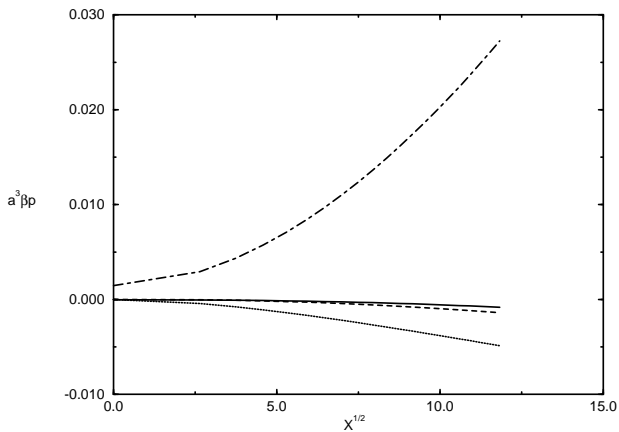}
\end{center}
\vspace*{-1.0cm}
\begin{minipage}{0.48\textwidth}
\caption{Relative contributions to the osmotic pressure inside a polyelectrolyte solution ($\rho_{salt}=10^{-3}M$, see caption to figure $6$ for details) in order of importance: mixing (dot-dashed line); polyion-counterion (dotted line); polyion-polyion (dashed line); counterion-coion (solid line).}
\label{Fig.7}
\end{minipage}
\end{figure}

\begin{figure}[h]

\begin{center}
\epsfxsize=8.cm
\leavevmode\epsfbox[40 120 550 520]{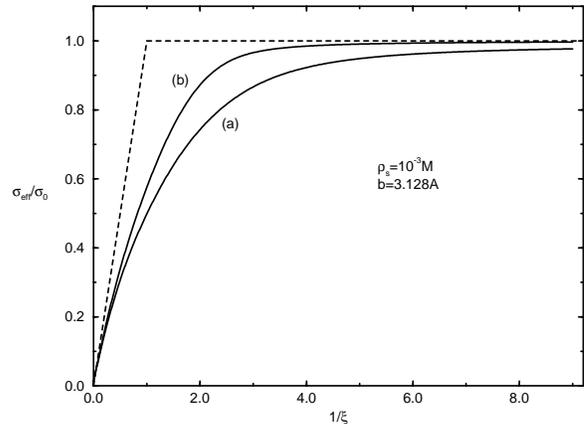}
\end{center}
\vspace*{-0.5cm}
\begin{minipage}{0.48\textwidth}
\caption{Dependence of the effective charge density $\sigma_{eff}/\sigma_0$ on the inverse Manning parameter $1/\xi$. Two densities of polyions are considered: (a) $\rho_p=10^{-4}$M; (b) $\rho_p=10^{-14}$M. The dashed line is the Manning's infinite dilution theory. Note that at infinite dilution there is no counterion association for $\xi<1$, and the transition appears sharply at $\xi = 1$, with $\sigma_{eff}/\sigma_0=1-1/\xi$ (see text in Conclusion).}
\label{Fig.8}
\end{minipage}
\end{figure}

\begin{figure}[h]

\begin{center}
\epsfxsize=8.cm
\leavevmode\epsfbox[40 120 550 520]{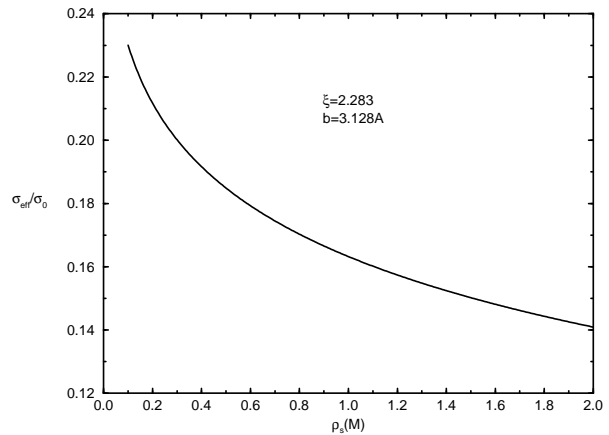}
\end{center}
\begin{minipage}{0.48\textwidth}
\caption{Dependence of the relative effective charge density $\sigma_{eff}/\sigma_0$ on the density of salt $\rho_{salt}$. The density of polyion is fixed: $\rho_p=10^{-8}$M.}
\label{Fig.9}
\end{minipage}
\end{figure}

\section*{\it ACKNOWLEDGMENTS}

P.K. would like to thank Dr. M\'ario  N. Tamashiro for the help with numerics. Y.L. would like to acknowledge helpful conversations with M.E. Fisher and G. Stell. This work was supported in part by CNPq - Conselho Nacional de
Desenvolvimento Cient\'{\i}fico e Tecnol\'ogico and FINEP -
Financiadora de Estudos e Projetos, Brazil.

\end{multicols}

\begin{thebibliography}{40}
\bibitem {obs1} In Ref.[$6$] the second term of eq.(\ref{e15}) was neglected, since it is of higher order in density and will not affect the infinite-dilution limit. However, it will produce a contribution as soon the density becomes non-vanishing. In this case the value of the parameter $c(T^*)$ defined in Ref.[$6$] should be modified by an addition of $1$, becoming $c(T^*)=2 \gamma _E +1/2 + \ln (\pi/T^*)$.
\end{thebibliography}
\end{document}